\documentstyle[12pt]{article}
\begin{document}
\begin{center}
{\ {\Large {\bf The structure of a source from interferometry   }}}\\

\vskip 0.4cm

{\ {\Large {\bf measurements in heavy ion collisions }}}\\

\vskip 0.5cm

{\sl Peter Filip}\\
\vskip 0.25cm
Institute of Physics, Slovak Academy of Sciences \\
842 28 Bratislava, Slovak Republic \\
Department of Physics, Comenius University \\
842 15 Bratislava, Slovak Republic \\
E-mail: filip@savba.sk \\

\vspace{0.32cm}
\end{center}

\begin{abstract}
Some experimental results of correlation functions in Bose-Einstein (B-E)
interferometry measurements exhibit a non smooth behaviour - oscillations.
Possible origin of such a behaviour in non-trivial spatial distribution of
the source is considered here.
A method for obtaining additional information about the structure of
source emitting identical particles is suggested.
It is shown that this information could be extracted by a one dimensional
fourier analysis of the filtered correlation function. Further possible
enhancement of the method is sketched.
Some technical aspects of the proposed technique
are discussed.

\end{abstract}
\vspace{0.2cm}
\begin{center}
{\bf Part1: Introduction}
\end{center}

\vspace{0.4cm}

The idea of measuring properties of a source of identical particles by
correlation experiments originated in radioastronomy. The angular diameter
of the star Sirius was successfully determined by HBT method [1] in
agreement with predictions. In experiments on antiproton-proton annihilation
in 1959 the correlation effect in like-sign two pion angular distribution
was discovered. This effect was interpreted in [2] as a result of B-E
interference of the wave functions of emitted pions. In the simplest,
sufficient for our purposes form the correlation function for a static
source can be up to the normalization factor expressed as:
\begin{equation}
C(\Delta \vec p) \simeq \int \int f(\vec x_1)f(\vec x_2)
\vert \Psi _{12}(\vec x_1,\vec x_2,\vec p_1,\vec p_2) \vert ^2
d^3 \vec x_1 d^3 \vec x_2
\label{eq1}
\end{equation}
where $\vec p_1, \vec p_2 $ are momenta of emitted identical bosons and
$f(\vec x) $ describes the geometrical distribution of the source. The squared
absolute value of $\Psi _{12} $ is:
\begin{equation}
\vert \Psi _{12} \vert ^2=
{1 \over 2} \vert e^{i\vec p_1 \vec x_1} e^{i\vec p_2 \vec x_2}
+e^{i\vec p_1 \vec x_2} e^{i\vec p_2 \vec x_1} \vert ^2
=1+\cos (\Delta \vec x \cdot \Delta \vec p)
\label{eq2}
\end{equation}
and it depends only on the relative momentum of the emitted bosons
$\Delta \vec p =\vec p_1 - \vec p_2 $ and on the relative position
of the emission points $\Delta \vec x = \vec x_1 - \vec x_2 $. For a
Gaussian source $f(\vec x)\simeq e^{-(x^2/R^2)} $ the correlation function (1)
can be calculated analytically:

\begin{equation}
C(\Delta \vec p) \simeq
\int \int \vert \Psi _{12} \vert ^2 e^{-(\vec x_1^2 - \vec x_2^2)/R^2}
d^3 \vec x_1 d^3 \vec x_2 =
1+e^{-\vert \vec p_1 - \vec p_2 \vert ^2 R^2 /2}
\label{eq3}
\end{equation}

By fitting experimental data for the correlation function with (3) an
approximate radius of the spherical source can be determined.
A more complicated Gaussian parametrization of the correlation function
 [3] separates the longitudinal, outward and sideward dimensions of the source.
Statistical errors of correlation function data in heavy ion collisions (HIC)
experiments became smaller but the interesting structure of the correlation
function in the higher relative momenta region is still present in the recent
results [4]. The influence of Coulomb final state interaction leads to
the more interesting shape
of the theoretical correlation function for the like sign charged particles
[5,6], but it is not rich enough to explain any oscillations of the
experimental correlation function in the
$\Delta p > 100 MeV/c $ region\footnote
{ The oscillations in theoretical shapes of the correlation
functions for photons can be found in the paper [7] (but see also [9]) }.
Perhaps, it could
 be possible to reveal some information about the source from this behaviour
 of the correlation function.

In the following section we define the longitudinal correlation function which
is subsequently used in the definition of distance spectrum of the source in
Sec.3. In Sec.4 we discuss the reconstruction of the distance spectrum
function from the longitudinal correlation function and some technical
aspects of this procedure. In Sec.5. we present a simple example as
the illustration of the method. In the last part we consider the possibility
to apply a tomography techniques based on the approach presented here.

In the whole paper we deal only with the influence of the spatial
distribution of source on the correlation function. The effects of the
final state interactions and the coherence of emission are not considered
here. It is assumed that they do not affect the principles of the method
presented.

\begin{center}
{\bf Part2: Longitudinal Correlation Function}
\end{center}

Let us consider a spatial static three-dimensional source emitting pairs of
boson particles. Because of B-E interference effect, the emission amplitude
of a pair of particles with relative momentum
$\Delta \vec p =\vec p_2 - \vec p_2$
depends on the relative position of the emission points. We can decompose the
relative momentum of the detected pair into transverse and longitudinal
components $\Delta \vec p = \Delta \vec p_t + \Delta \vec p_l $, where the
longitudinal direction is chosen along the beam in the HIC experiment.
Thus we have:
\begin{equation}
\Delta \vec p \cdot \Delta \vec x =
(\Delta \vec p_l + \Delta \vec p_t )\cdot
(\Delta \vec x_l + \Delta \vec x_t ) =
\Delta \vec p_l \cdot \Delta \vec x_l +
\Delta \vec p_t \cdot \Delta \vec x_t
\label{eq4}
\end{equation}
Let us choose events with a small transverse component of relative momentum

\begin{equation}
\vert \Delta \vec p_t \vert <<
\vert \Delta \vec p_l \vert
\label{eq5}
\end{equation}

Then the transverse separation of emission points does not have influence
on the correlation function for our set of data. An analogous consideration
can be made for the set of events filtered by the condition
\begin{equation}
\vert \Delta \vec p_t \vert >>
\vert \Delta \vec p_l \vert
\label{eq6}
\end{equation}
or one can choose the sideward and outward selections used e.g. in [3] but we
shall now concentrate on the correlation function for data events fulfilling
the condition (5) - the longitudinal correlation function. In this case after
the substitution (4) into (1) and assuming (5) we have:
\begin{eqnarray}
C(\Delta \vec p)&=&\int \int f(\vec x_1)f(\vec x_2)
(1+\cos (\Delta \vec x \cdot \Delta \vec p)) d^3 \vec x_1 d^3 \vec x_2 =
                                                                 \nonumber \\
&=&\int \int F(x^1_l)F(x^2_l)(1+\cos (\Delta x_l \cdot \Delta p_l ))
dx^1_l dx^2_l
\label{eq7}
\end{eqnarray}
where $F(x_l)=\int f(x_l,\vec x_t)d^2 \vec x_t $.

This means, that our 3$D$ source produces
$ C(\Delta p_l ) $
like a one dimensional source with the distribution
$F(x_l)$.

\begin{center}
{\bf Part3: Distance Spectrum of the Source}
\end{center}

Let us find how the longitudinal correlation function
$C(\Delta p_l)$
looks like for a two-point source (see Fig.1a). This two point source has a
distribution function
$F(x)=1/2 (\delta (x-a)+\delta (x-b)) $.
The longitudinal correlation function then is:
\begin{equation}
C2(d,\Delta p_l)=\int F(x_1)F(x_2)(1+\cos (\Delta p_l (x_1 - x_2 )))dx_1 dx_2
=1+ \cos (\Delta p_l \cdot d)
\label{eq8}
\end{equation}
where $d=\vert a-b \vert $ is the separation of emission points.
For a source with the distribution function
$F(x)=P_1 \cdot \delta (x-x_1) + P_2 \cdot \delta (x-x_2) +
P_3 \cdot \delta (x-x_3) $ (Fig.1b), where $P_1 + P_2 + P_3 = 1 $ the result
is:
\begin{equation}
C(\Delta p_l) = \Theta +
P_{12} \cdot C2(d_{12}, \Delta p_l) +
P_{13} \cdot C2(d_{13}, \Delta p_l) +
P_{23} \cdot C2(d_{23}, \Delta p_l)
\label{eq9}
\end{equation}
where $P_{ij}=P_i \cdot P_j $, $d_{ij}=\vert x_i - x_j \vert $ and $\Theta $
is a constant shift which will not be important in our further calculations.
Based on this it seems appropriate to define the Distance Spectrum of a source.
Each source emitting pairs of particles has its 'squeeze' - Distance Spectrum.
It is the probability distribution for the emission of the pair in the distance
$D$ of emission points. For the two-point source the distance spectrum is
$S(D) \simeq \delta (D-d_0)$ (see Fig.1a),
for our three-point source (8) the distance spectrum is
$S(D) \simeq P_{12} \delta (D - d_{12})
+ P_{13} \delta (D - d_{13})
+ P_{23} \delta (D - d_{23})$ (see Fig.1b). In general the distance spectrum
for a one dimensional static source can be expressed as:

\begin{equation}
S(D)=\int f(x_1) f(x_2) \delta (D-\vert x_1 - x_2 \vert )
d x_1 d x_2
\label{eq10}
\end{equation}

Now we can write the longitudinal correlation function produced by a source
with distance spectrum $S(D)$:
\begin{equation}
C(\Delta p_l) = \Theta ' + \int S(D) C2(D,\Delta p_l)dD
\label{eq11}
\end{equation}

A distance spectrum does not contain complete information about the spatial
distribution of a source. Slightly different spatially distributed sources can
have the same $S(D)$ (see also Sec.5.). The distance spectrum of another
point-like source is shown in Fig.1c .

\begin{center}
{\bf Part4: Inverse Transformation}
\end{center}

Expression (1) for the correlation function can be after the formal
integration written in the form:
\begin{equation}
C(\Delta \vec p)=1+\vert \tilde f (\Delta \vec p) \vert ^2
\label{eq12}
\end{equation}
where $\tilde f (\Delta \vec p)=\int e^{i\Delta \vec p \cdot \vec x}f(\vec x)
d^3 \vec x $. The absolute value in expression (12) destroys the phase
information in the fourier picture $\tilde f (\Delta \vec p) $ of source
distribution. This breaks the possibility
to perform the inverse fourier transformation in order to get $f(\vec x)$.

In this section we shall show that the remaining information in
$\vert \tilde f (\Delta \vec p)\vert $ has certain physical meaning and
that its extraction is at least theoretically possible.

The form of expression (11) is sufficiently simple. Therefore one could
consider a possibility to gain the function $S(D)$ from $C(\Delta p_l)$.
As we shall see, the $S(D)$ is just the information contained in
$C(\Delta p_l)$.
According to (11) and (8) we have

\begin{equation}
C(\Delta p_l) = \Theta '' + \int S(D) \cos (\Delta p_l \cdot D)dD
\label{eq13}
\end{equation}

The constant shift factor can be separated from the correlation function data
and so we can write:
\begin{equation}
\tilde {C}(\Delta p_l) = \int_0^{\infty} S(D) \cos (\Delta p_l \cdot D)dD
\label{eq14}
\end{equation}
where $\tilde {C}(\Delta p_l)$ is the correlation function with the shift
factor removed. Expression (14) is fourier transformation and we can try to
perform the inverse fourier transformation in order to obtain $S(D)$
from experimental data.

\begin{equation}
S(D) \simeq \int_0^{\infty } \tilde {C}(\Delta p_l)
\cos(\Delta p_l \cdot D) d(\Delta p_l)
\label{eq15}
\end{equation}

Because of the fact, that we measure $C(\Delta p_l)$ only in discrete points
it is necessary to replace integration by summation.

\begin{equation}
S_F(D)\simeq \sum_i^{\hbox {\tiny {points}}} \tilde {C}(\Delta p_l^i)
\cos (\Delta p_l^i \cdot D)(\Delta p_l^{i-1} - \Delta p_l^i )
\label{eq16}
\end{equation}

However a more relevant problem is that the integration region in (15)
includes the values of $\Delta p_l$ which cannot be measured
( high $\Delta p_l$ ). The cut-off in $\Delta p_l$ is present
due to energy conservation considerations and due to the growth of statistical
errors in a high $\Delta p_l$ region. The influence of such a cut-off on
the result of Fourier transformation leads to the oscillations in the resulting
$S(D)$ (see Fig.2a) regardless of the number of measured points
in the region $(0,\Delta p_{\hbox {\tiny {max}}})$. Similar problems occur
e.g. in
digital FIR\footnote {FIR = finite duration impulse response } filter design.
Fortunately these oscillations can be suppressed by the
Method of Windows [8]. The method is based on the simple multiplication of
the original function to be fourier transformed by the Window Function,
which suppresses the values of the original function near the cut-off
(see Fig.2b).

\begin{equation}
S_W (D) \simeq \sum_i^{\hbox {\tiny {points}}} W(\Delta p_l^i)
\tilde {C}(\Delta p_l^i)
\cos (\Delta p_l^i \cdot D)(\Delta p_l^{i-1} - \Delta p_l^i )
\label{eq17}
\end{equation}

There are several types of the Window Function. In Fig. 2b a simple Gaussian
function is used. The amplitude of the oscillations depends also on the shape
of the exact (without cut-off) result of the Fourier transformation. The
exact result of Fourier transformation of the function in Fig. 2a without
cut-off would be Dirac delta function at the point $d_0 = 3$fm.
Such a shape leads to a big oscillations.

The influence of statistical errors of correlation function data to the
resulting distance spectrum which is crucial for the eligibility of the
method is not considered in this paper.

\begin{center}
{\bf Part5: Structured Time Dependent Source}
\end{center}

As we have already mentioned the sources with the same $S(D)$ can be different,
but in spite of that, having the distance spectrum of a source we can make
some conclusions about the source. At least the average radius
\begin{equation}
\bar R = \int D\cdot S(D) dD
\label{eq18}
\end{equation}
and maximal radius $R_{\hbox {\tiny {max}}} = D_{\hbox {\tiny {max}}}$ in
$S(D)$ can be found from $S(D)$.

Let us consider a three-dimensional source consisting of two separated regions
(Fig.3c). Both regions $A$ and $B$ contribute to distance spectrum of a source
in
longitudinal (5) analysis in the interval  $0 < D < R$. When one of the
particles comes from A and the second from B they can be emitted in a
distance interval $K-R < D < K+R$. The distance spectrum $S(D)$ in the interval
$R < D < K-R$ is suppressed as much as the parts $A$ and $B$ are located.
Original spectra $S_{MC}(D)$ for such a source computed by Monte-Carlo
simulation, the correlation functions calculated according expression (11) at
200 points in the $\Delta p_l < 2$GeV interval and the results of
inverse transformation (16) for different separations $K$ of parts $A$, $B$ of
the source are in figures 3a, 3b, 3c.

The real experimental situation is however a time dependent distribution
of the source. The general expression for the correlation function
in this case is much more complicated than expression (1):

\begin{equation}
C(\Delta p)=\int \int
w\bigl (x_1, {{p_1 + p_2}\over {2}} \bigr )
w\bigl (x_2, {{p_2 + p_1}\over {2}} \bigr )
(1+\cos(\Delta x^\mu \Delta p_\mu)) d^4x_1d^4x_2
\label{eq19}
\end{equation}

However the influence of the spatial structure of the source could still
demonstrate itself in the oscillations of the correlation function at
the higher relative momentum region.

Following the simple approach based on (1)
in the case of the time evolution we have to take into consideration
that the bosons emitted in different times do interfere.
This means, that the same region of the hadron gas or quark gluon plasma
emitting the bosons in
two different times $t_1$ and $t_2$ at the positions $\vec x_1$ and $\vec x_2$
contributes to our distance spectrum
of the source like a two different regions emitting the bosons at the same time
but in separate points
$\vec x_1$ and $\vec x_2$. Therefore an important question arises: How big can
be the time difference of the processes of emission for interfering bosons ?

Let us denote the duration of the process of emission
of a single  boson as $\tau _0$.
As an upper limit for $\tau _0$ the formation time of the emitted boson
which is of the order of 1fm/c [11] could be chosen.
It seems to be
reasonable to require a time overlapping of the processes of emission for the
interfering bosons in the B-E interference phenomenon.
Assuming the Lorentz dilatation factor $\gamma $ for $\tau _0$ of  emitted
boson we obtain the
rough restriction on the time difference $\Delta t$
of emissions of the interfering bosons
in the form:
\begin{equation}
\Delta t < \tau _0 \cdot \gamma
\label{eq20}
\end{equation}

For a typical momentum of pions produced in heavy ion collisions the time
$\Delta t$
is long compared to $m_{\pi}^{-1}$ what is necessary for the influence of
B-E interference on  correlation function [10].

Different spatial distributions in a different time
intervals lead to different distance spectra $s(D,t)$. In our simple approach
the correlation function for a time dependent source could be expressed for
$\tau _0 \rightarrow 0$ as:
\begin{equation}
\tilde {C}(\Delta p_l) = \int \int_0^{\infty} s(D,t)
\cos (\Delta p_l \cdot D)dDdt =
\int_0^{\infty} \biggl \lbrack \int s(D,t) dt \biggr \rbrack
\cos (\Delta p_l \cdot D ) dD
\label{eq21}
\end{equation}
but for $\tau _0 \neq 0$ the information about the $s(D,t)$ in the correlation
function is smudged with the uncertainty proportional to $\Delta t$ given
by (20).

It is clear, that by the inverse transformation (15) we can obtain only $S(D)$,
the result of the time integration in (21). A combination of the results
obtained for different kinds of identical particles emitted in different stages
of the evolution of the heavy ion collision or the results obtained for
filtered events which can be produced only
during characteristic and short time intervals of the evolution process
could bring some information about $s(D,t)$.

\begin{center}
{\bf Part6: B-E Tomography}
\end{center}

In this section we shall deal with the set of events fulfilling condition (6)
- transversal events. Analogously to Sec.2. the longitudinal separation of
emission points does not have influence on the transversal correlation
function\footnote{ the correlation function for the transversal set of events}
now, but condition (6) is not so selective as condition (5). We have a freedom
in azimuthal direction of $\Delta \vec p_t$ and the emission points lie in the
plane orthogonal to the beam direction (see Fig.4). It is possible to apply
some  further condition to have the same one-dimensional situation like in the
preceding sections. For example the condition
\begin{equation}
\Delta \vec p_t \cdot \vec n_\varphi >>
\vert \Delta \vec p_t \times \vec n_\varphi \vert
\label{eq22}
\end{equation}
where $\vec n_\varphi $ is a normalized vector orthogonal to the beam direction
(see Fig.4),
selects
the pairs with relative momentum in the direction of vector $\vec n_\varphi$.

However the cylindrical symmetry of the radiating volume in HIC, at least for
the central collisions, instigates us to try to gain the whole two-dimensional
information from $C(\Delta \vec p_t )$.
Let us define two-dimensional distance spectrum as:
\begin{equation}
S(\vec D)=
\int \int f(\vec x_1)f(\vec x_2)\delta (\vec D - (\vec x_1 - \vec x_2 ))
d^2 \vec x_1 d^2 \vec x_2
\label{eq23}
\end{equation}
$S(\vec D)$ is the probability distribution of emission of the pair of
particles from the points at relative position $\vec x_1 - \vec x_2 =\vec D$
on the transversal plane. Then for our transversal correlation function
\begin{equation}
\tilde C(\Delta \vec p_t) = \int f(\vec x_1)f(\vec x_2)
\cos (\Delta \vec p_t \cdot \Delta \vec x) d^2\vec x_1 d^2\vec x_2
\label{eq24}
\end{equation}
we can write
\begin{equation}
\tilde C(\Delta \vec p_t) = \int S(\vec D)\cdot
\cos (\Delta \vec p_t \cdot  \vec D) d^2 \vec D
\label{eq25}
\end{equation}
what can be proven by inserting (23) into (25).

For a set of transversal events fulfilling the condition (22) specified by the
angle $\varphi$ the correlation function is:
\begin{eqnarray}
\tilde C(\Delta p_{\varphi} ) &=&
\int \int S(\vec D)
\cos (\Delta \vec p_\varphi
      \cdot
      (\vec D_\varphi + \vec D^\perp _\varphi )
d D_\varphi
d D^\perp _\varphi = \nonumber \\
&=&\int \left[ \int S(\vec D) d D^\perp _\varphi \right]
\cos (\Delta p_\varphi \cdot D_\varphi )
d D_\varphi = \\
&=& \int S_\varphi (D_\varphi ) \cdot \cos
    (\Delta p_\varphi \cdot D_\varphi )
d D_\varphi \nonumber
\label{eq26}
\end{eqnarray}
where the index "$\ ^\perp _\varphi$" describes the orthogonal and the index
"$\ _\varphi $"  parallel direction
to the vector $\vec n $.
Thus by the inverse transformation (15) we can
obtain the projection $S_\varphi (D)$ of the two-dimensional distance spectrum
$S(\vec D)$ to the direction determined by the angle $\varphi$. This can be
done for any direction, for example in $2^\circ $ steps of our
angle $\varphi $.

 From such a set of projections one can get the whole $S(\vec D)$ by
Radon Transformation which is used in tomography [13].

It is hard to imagine
that this technique could be applied to the interferometry experimental data
in heavy ion collisions. However in the future HIC experiments e.g. at LHC the
multiciplities/event could allow to do something close to the ideas
presented here.  For the final
decision whether this tomography method is eligible or not a careful and
precise estimate of statistical errors is necessary. Both the subjects - the
description of Radon Transformation and the analysis of statistical errors
are not included in this paper.

One can imagine that the proposed two-dimensional tomographic method
is applicable and even more, that a time dependent three-dimensional
distance spectrum can be somehow gained. The question arises: What could we
see from $S(\vec D)$ or $S(\vec D,t)$ ?
Author thinks that it could be
possible to see some signatures of the phase transition in this kind of
analysis.
 In cosmology the phase transition as the only explanation of the
inhomogeneities in the distribution of matter in universe is used.
We can hope that the phase
transition in HIC could lead to the formation of the inhomogeneities - the
bubbles in the volume of the collision. These bubbles emitting particles
in a different strength
than the surrounding medium could be seen in our distance spectra.

\begin{center}
{\bf Part7: Summary}
\end{center}

We have studied the influence of the spatial distribution of a source on the
correlation function in a set of longitudinal (5) events. It was found,
that spatially complicated structure of the source can lead to oscillations
of the correlation function.
It is shown that from the longitudinal correlation function the distance
spectrum of a source in the longitudinal direction can be derived. The radius
of the source, which is used to characterize the longitudinal size
of the
collision volume is
expressed as a simple integral of the distance spectrum.
Some technical aspects of the suggested method are discussed. The problem of
the time dependence of spatial distribution of a source is also considered.
An analogous method for transversal set of events and the possibility to use
the tomography method in this approach are considered.
The proposed technique could evoke some ideas how to enhance the interferometry
methods in the future heavy ion collisions experiments.

\begin{center}
{\bf Part8: Acknowledgements}
\end{center}

Author would like to express special thanks to Univ. Doz. M.Faber and
M.Meinhart for the precious discussions during the work on the paper.
The help of A.Nogov\'a, \v S.Olejn\'{\i}k, prof. J.Pi\v s\' ut and
the discussions with the scientists from the Institute of Measurements
of Slovak Academy of Sciences were also
valuable. The computer calculations were performed on the computers of
Technische Universit\"at Wien. Considerable part of this work was supported
by Bundesministerium f\"ur Wissenschaft und Forschung stipendium
ZL. AD/1021-5/1993 Vienna, Austria.

\newpage

\newpage
\section*{Figure captions}

\begin{description}

\item{[Fig.~1a,1b,1c]}
The point-like sources and their distance spectra obtained by
Monte-Carlo simulation in the agreement with the analytical probability
calculation.

\item{[Fig.~2a,2b]}
The correlation function $C2(\Delta p)$ for two-point source
(distance $d_0 = 3\hbox {fm} $) and the result of the inverse transformation
without (2a) and with (2b) the Method of windows.

\item{[Fig.~3a,3b,3c]}
The distance spectrum $S_{MC}(D)$ calculated by Monte-Carlo
simulation, the corresponding correlation function $\tilde {C2} (\Delta p)$
at 200 points and a result of inverse transformation (15) for the
three different separations of the sources $A$, $B$.

\item{[Fig.~4]}
The plane transversal to the beam direction and the vector $\vec n $ with
direction determined by the angle $\varphi $. For the central collision
the zero angle direction can be chosen randomly.  For a non central collision
an asymmetry in azimuthal direction in the distributions of the produced
particles [12] could be used to determine the zero angle direction.

\end{description}


\begin{thebibliography}{x}

\bibitem{1}
{\sc H.Brown, R.Q.Twiss,} {\em	Nature} {\bf 178} 1046 (1956)

\bibitem{2}
{\sc G.Goldhaber et al.} {\em  Phys. Rev.} {\bf 120 N1} 300 (1960)

\bibitem{3}
{\sc M.Herrmann, G.F.Bertsch,} {\em DOE/ER/40561-141-INT94-00-57}
(hep-ph/9405373)

\bibitem{4}
{\sc Hans Boggild,} {\em NA 44 data preliminary}  presented at the European
Research Conference on Nuclear Physics, Vuosaari Finland, 17 - 22 June 1994

\bibitem{5}
{\sc M.G.Bowler,} {\em Z.Phys.C-Particles and Fields} {\bf 39} 81-88 (1988)

\bibitem{6}
{\sc M.G.Bowler,} {\em	Phys.Letters B} {\bf  270} 69-74  (1991)

\bibitem{7}
{\sc K.Srivastava, J.I.Kapusta,} {\em Phys. Review C} {\bf 48, N3} 1335 (1993)

\bibitem{8}
{\sc A.V.Oppenheim, R.W.Schafer,} {\em Digital Signal Processing} Prentice
Hall, Englewood Chiffs,N.J.,1975

\bibitem{9}
{\sc A.Timmerman et.al.} {\em Photon Interferometry of Quark Gluon Dynamics
Revisited} hep-ph/9405232

\bibitem{10}
{\sc M.Gyulassy et.al. } {\em Physical Review C } {\bf 20, N6}  2267 (1979)

\bibitem{11}
{\sc J.Pi\v s\' ut et.al.} {\em Formation Time of Hadrons and Density of
Matter Produced in Relativistic Heavy Ion Collisions} Comenius University
Bratislava Preprint {\bf Ph2-94}

\bibitem{12}
{\sc S.Voloshin and Y.Zhang,} {\em Flow Study in Relativistic Nuclear
Collisions by Fourier Expansion of Azimuthal Particle Distributions }
to be published

\bibitem{13}
{\sc G.T.Herman,} {\em Image Reconstruction from Projections - The Fundamentals
 of Computerized Tomography}  Academic Press, New York 1980

\end{thebibliography}
\end{document}